\documentclass[preprint,authoryear,12pt]{elsarticle}
\usepackage{amssymb}
\usepackage{epsfig}
\usepackage{graphicx}
\usepackage{multirow}
\usepackage{url}
\biboptions{square}
\journal{Journal of Informetrics}
\begin{document}
\begin{frontmatter}
\title{On the  time dependence of the $h$-index}
\author{Riccardo Mannella}\ead{mannella@df.unipi.it}
\author{Paolo Rossi\corref{ca}}\ead{rossi@df.unipi.it}
\address{Dipartimento di Fisica "Enrico Fermi", Universit\`a di Pisa, 56127 Pisa, Italy}
\cortext[ca]{Corresponding author}

\begin{abstract}
The time dependence of the $h$-index is analyzed by considering the average behaviour of $h$ as a function of the academic age $A_A$ for about 1400 Italian physicists, with career lengths spanning from 3 to 46 years. The individual $h$-index is strongly correlated with the square root of the total citations $N_C$: $h \approx 0.53 \sqrt{N_C}$. For academic ages ranging from 12 to 24 years, the distribution of the time scaled index $h/\sqrt{A_A}$  is approximately time-independent and it is well described by the Gompertz function. The 
time scaled index  
$h/\sqrt{A_A}$ has an average approximately equal to 3.8 and a standard deviation approximately equal to 1.6. Finally, the time 
scaled index $h/\sqrt{A_A}$ appears to be strongly correlated with the contemporary $h$-index $h_c$.
\end{abstract}
\end{frontmatter}

\section{Introduction}

One of the purposes of modern bibliometrics is to introduce some quantitative indicators of the scientific production of individuals, aiming at establishing some rough classification or ranking. An indicator which has been gaining much attention is the Hirsch index 
$h$~\citep{Hirsch:2005}: given an individual with $N$ publications, $h$ is defined as the number of papers which received at least $h$ citations, while the remaining $N-h$ papers received less than $h$ citations. Given that the $h$-index increases monotonously with the age of the scientist involved, its time dependence has been a relatively long-standing problem of bibliometrics, with deep consequences on the possibility of comparing scientists showing substantial differences in their academic age $A_A$, that is the length (in years) of their academic career.

In his original paper, \citet{Hirsch:2005} proposed that the $h$-index would be growing roughly linearly in time, and he therefore suggested the introduction of the 
$m$-index, simply defined as the ratio between the $h$-index and the time lapse $T_L$ (in years) between the first publication and the present date: $m=h/T_L$. 

However \citet*{GunsRousseau:2009} showed by numerical simulations in a model of the citation dynamics that  the functional dependence of the growth may be affected by a number of different deterministic and stochastic factors, and linearity is not always assured. Absence of linearity was observed also by \citet{Egghe:2009a,Egghe:2009b,Egghe:2010} and \citet*{Wu:2011}. In view of these results, it seems rather difficult to construct a robust indicator allowing a precise ranking of scientists with different career lengths.

On the other hand, when the goal is to establish a benchmark of scientific quality and productivity acting as a \textit{threshold} for recruitment and promotion, we are no longer bound to exploring the exact dependence of the index on individual careers: rather, we may consider the \textit{statistical average} for sufficiently large groups as a proxy  for an “ideal” temporal dependence of scientific production and of its impact, and to establish whether these averages show some general behaviour.

We collected the bibliometric data of about 1400 Italian physicists (randomly chosen among the approximately 2400 Physics academic staff employed in Italian Universities at the end of 2010) using the SCOPUS database, grouped according to the date of their first scientific publication appearing on the database, from years 1965 to 2008. We then computed the average of the total citations and of the $h$-index for each annual group, and studied the correlations of these indicators between each other and with time. Clearly, the $T_L$ for each group is given by the difference between the time of data extraction and the year labeling each annual class, and we identified the academic age $A_A$ with $T_L$.

We anticipate our main conclusions:\begin{itemize}
\item the individual $h$-index is very strongly correlated with the total number of individual citations, as suggested by \citet{Hirsch:2005} and emphasized by \citet{Nielsen:2008}.
\item the ratio between (group averaged) total citations and academic age shows three markedly different behaviours. The ratio grows (roughly linearly) with time during the first ten years; it  stabilizes at a relatively constant (plateau) value for at least fifteen years; it then decreases to reach a second constant, but lower, value, for longer academic ages $A_A$
\item a similar pattern (which we believe to follow from the observed time dependence of the above ratio) is shown by the ratio between the (group averaged) $h$-index and the \textit{square root} of the academic age $A_A$
\item the ratio between the individual $h$-index and the \textit{square root} of the academic age ($h/\sqrt{A_A}$) appears to be strongly correlated with the contemporary $h$-index $h_c$
\item to assess scientists who have been active for more than ten years, it appears reasonable to compare the index $h/\sqrt{A_A}$ to the observed plateau values
\end{itemize}

\section{The correlation between the total number of citations and the $h$-index}

As first explained by Hirsch, the relationship between total number of citations $N_C$ of individuals and their $h$-index is expected to take the general form  $N_C= a h^2$, with  $3<a<5$, although there seems to be no obvious theoretical reason why the parameter $a$ should have some special and universal value.

\begin{figure}[hbt]
\includegraphics*[width=8 cm]{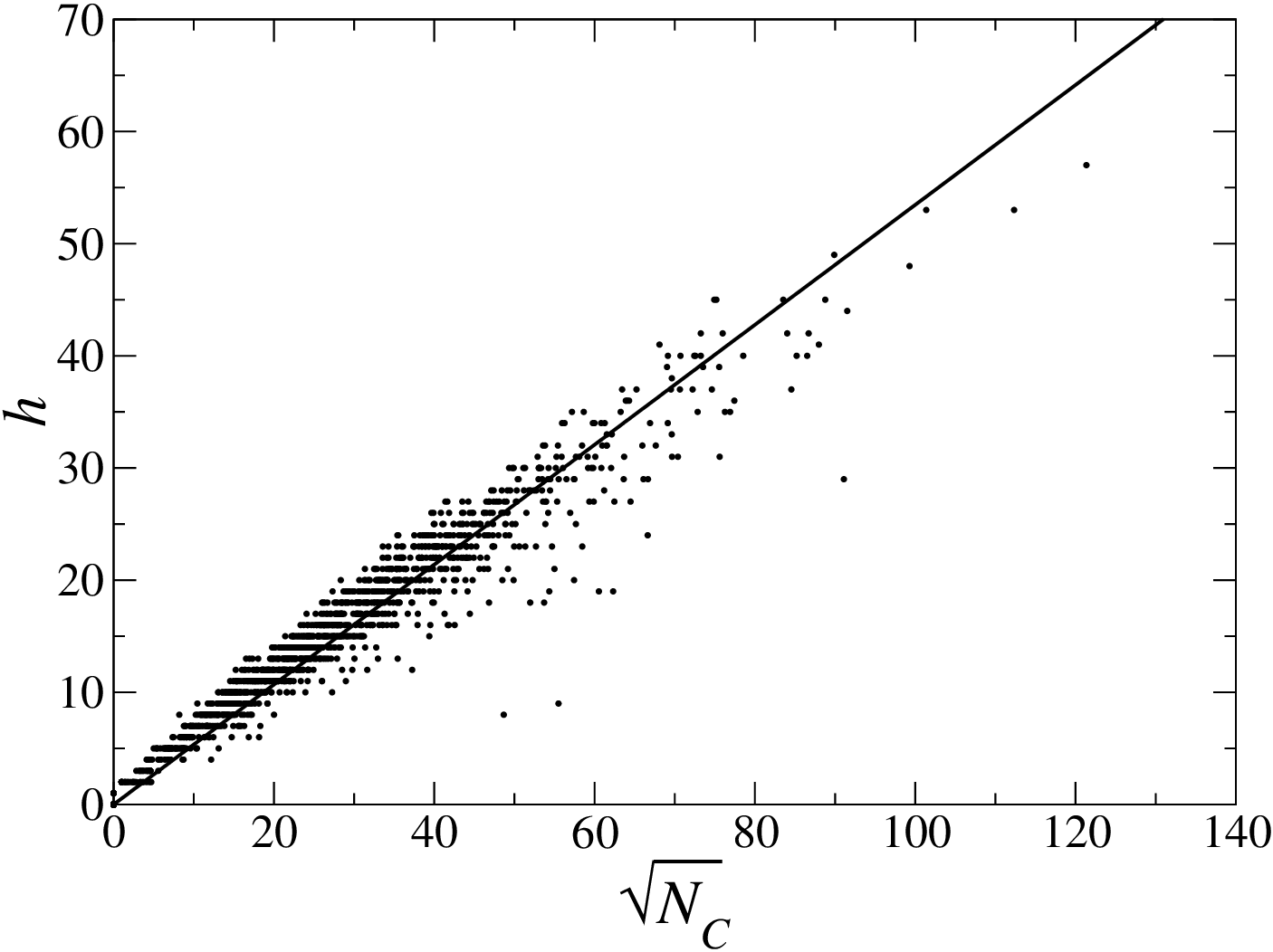}
\caption{The $h$-index vs the square root of the total number of citations $N_C$, for each scientist considered in this work (dots). The straight line is the relation $h = \alpha \sqrt{N_C}$, with $\alpha=0.53$ obtained through a best fit to the data}
\label{fig1}
\end{figure}

Fig.~\ref{fig1} shows the relation between $h$ and $N_C$: a clear linear relation is visible when $h$ is plotted against $\sqrt{N_C}$, confirming the empirical suggestion by Hirsch. The correlation between the two variables in the plot is 0.97. The straight line is a best fit, using a relation of the form $h = \alpha \sqrt{N_C}$, and the resulting slope is $\alpha= 0.53$, corresponding to a value $a \approx 3.5$.

We also examined more restricted communities (like theoreticians and experimentalists, or senior and junior researchers) finding typically that $a$ changes only very mildly among different communities. The resulting $a$'s are summarized on table~\ref{table1}.

\begin{table}[hbt]
	\centering
	\begin{tabular}{lcccccc|cc}\hline\hline
		 & & FullP & AssoP & Rese & Total &  & Sample & Plateau \\ \hline
		\multirow{2}{*}{Astro}& $a$ & 3.23 & 3.28 & 3.64 & 3.33 &  & 71 & 4.4 \\
		 & $corr$ & 0.974 & 0.975 & 0.963 & 0.972 &  & 32 & 2.9 \\\hline
		\multirow{2}{*}{HEP - exp} & $a$ & 3.64 & 3.33 & 3.92 & 3.60 &  & 110 & 4.1 \\
		 & $corr$ & 0.967 & 0.967 & 0.895 & 0.951 &  & 99 & 3.2 \\\hline
		\multirow{2}{*}{HEP - the} & $a$ & 3.67 & 3.07 & 3.14 & 3.42 & & 79 & 3.8 \\
		& $corr$ & 0.971 & 0.967 & 0.940 & 0.970 & & 88 & 2.9 \\\hline
		\multirow{2}{*}{Matter - exp}& $a$ & 3.49 & 3.42 & 3.26 & 3.43 & & 158 & 3.7 \\
		 & $corr$ & 0.972 & 0.976 & 0.920 & 0.969 & & 155 & 3.0 \\\hline
		\multirow{2}{*}{Matter - the}& $a$ & 3.97 & 3.47 & 3.42 & 3.73 &  & 64 & 3.9\\
		 & $corr$ & 0.957 & 0.970 & 0.953 & 0.968 & & 47 & 3.0 \\\hline
		\multirow{2}{*}{AppPhys}& $a$ & 3.45 & 3.09& 2.83 & 3.19 &  & 72 & 3.1 \\
		 & $corr$ & 0.934 & 0.964 & 0.969 & 0.960 &  & 52 & 2.3 \\\hline
		\multirow{2}{*}{All}& $a$ &   &  & & 3.5 &  & 554 & 3.8 \\
	            & $corr$ &   &  & & 0.97&  & 473 & 2.9 \\ \hline\hline
	\end{tabular}
\caption{Summary of the parameter $a$, and the corresponding measure of correlation $corr$, evaluated considering different Physics research fields (Astro: Astronomy and Astrophysics; HEP: High Energy Physics; Matter: Condensed Matter, Atomic, Molecular and Optical Physics; AppPhys: Applied Physics; exp: Experimentalists; the: Theoreticians) and academic career progress (FullP: Full professors; AssoP: Associate professors; Rese: Research associates). The columns labeled Sample and Plateau are relative to the time dependence of the $h$-index, and they will be discussed further down in this paper.}
\label{table1}
\end{table}
We notice that indeed most of the $a$ values are in the range 3 to 4, and that the correlation coefficient is very close to one, in most cases. Furthermore, it is interesting to note that $a$ tends to change little within each age category, possibly with the exception of the HEP experimental research associates (who show an $a$ significantly larger than the $a$ of the research associates of other fields): this is easily understood when we recall that research in this field typically involves large collaborations: hence research associates, even at relatively young academic ages, possibly show bibliometric indicators typical of older academic staff in the field.

In the spirit of our approach, aiming at defining some benchmarks and thresholds rather than individual rankings, our first preliminary conclusion is that the total number of citations $N_C$ is as good an indicator as the $h$-index itself: 
this implies that $\frac{\sqrt{N_C}}{2}$ is a quite reasonable proxy of the $h$-index~\citep{Nielsen:2008}.

\section{Time evolution}
\subsection{The time dependence of the total number of citations}

The individuals under consideration have academic ages $A_A$ ranging from 3 to 46 years, and the dimension of the corresponding “age” groups  ranges from 4 to 63 units. We discarded a (small) number of cases corresponding to age values outside the above mentioned range because of the scarce statistical significance of the corresponding samples, and looked at the behaviour of an indicator defined as the total number of citations divided by the academic age ($N_C/A_A$) as function of the academic age.

\begin{figure}[hbt]
\includegraphics*[width=8 cm]{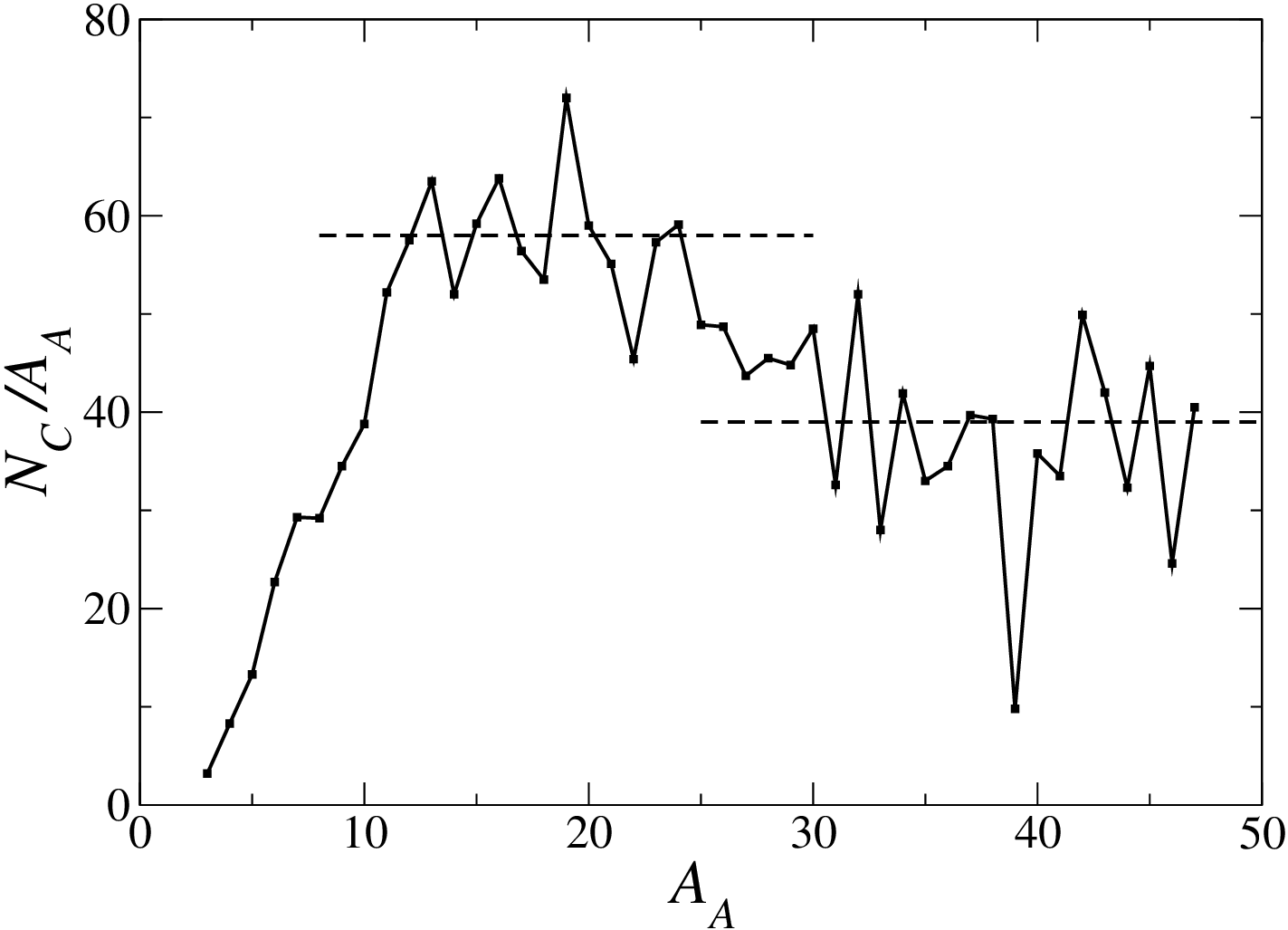}
\caption{The average number of citations divided by the academic age $N_C/A_A$ vs the academic age $A_A$ (solid broken line). The two dashed lines mark the two constant annual citation rates discussed in the text.}
\label{fig2}
\end{figure}

The result is shown in Fig.~\ref{fig2}. Despite some fluctuations mainly due to the small population of some age groups, three distinct time ranges characterized by different behaviours of the indicator are clearly visible:\begin{itemize}
\item In the academic age range between 3 and 12 years the indicator grows (roughly) linearly, starting from zero after a two-year time delay from the first publication date. Notice that a linear growth in the indicator would correspond to a quadratic growth in the total number of citations, and this is consistent with a (plausible) pattern of a constant publication rate and of a citation rate per publication staying constant for some years. Saturation occurs when older publications cease to be quoted and  the annual citation rate is kept constant only by the influx of new publications.
\item In the academic age range between 12 and 24 years the annual citation rate (barring fluctuations) stays constant at a (weighted) average value of approximately 58 citations per year. 
\item A rapid decline follows, and for academic ages above 30 years a new approximate stabilization occurs, with (weighted) values oscillating around 39 citations per year.
\end{itemize}

The decline in the annual citation rate might very well be explained by scientific “aging” occurring for members of the community, being typically in their sixties, and by a possible influence of a general growth in the number of citations observed in recent times, which tends to bias towards lower citation rate older researchers. We note, however, that the sharp decline and the subsequent lower level stabilization could be due to a possible bias present in our sample: approximately thirty years ago, Italian Universities underwent a massive permanent recruiting, and it is believed that not all the people recruited in those times (and who are still present in the system, and hence in our dataset) managed to keep productivity standards typical of more selected groups. 

\subsection{The time dependence of the $h$-index}

In view of the results presented in the previous sections, it is rather obvious to explore the behaviour of a time-normalized $h$-index obtained by taking the ratio between $h$ and the square root of the academic age $A_A$.

\begin{figure}[hbt]
\includegraphics*[width=8 cm]{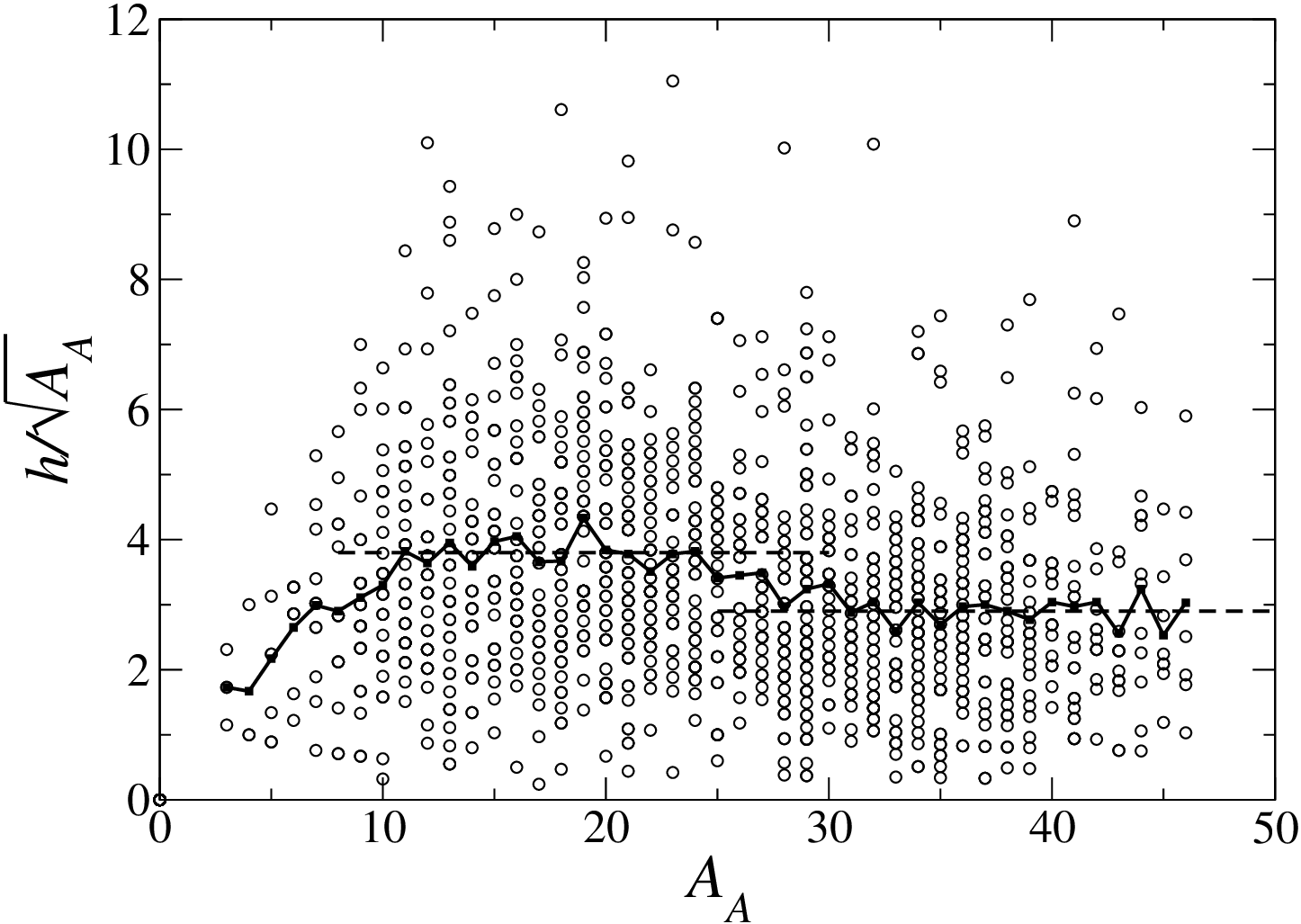}
\caption{The average $h$-index normalized to the square root of academic age $h/\sqrt{A_A}$ vs the academic age $A_A$ (black squares with a broken line). The two dashed lines mark the two constant values discussed in the text. Circles show the values of the normalized $h$-index of each researcher in our sample.}
\label{fig3}
\end{figure}

The result is summarized in Fig.~\ref{fig3}. Pleasantly enough, fluctuations are damped and the time pattern observed for the average number of citations (Fig.~\ref{fig2}) is even more evident. Following an initial growth, in the range between 12 and 24 years of academic age we observe a plateau value approximately equal to 3.8 (with a standard deviation approximately equal to 1.6), followed by a decline to a plateau value of 2.9 for academic ages larger than 30 years. The time dependence of $h\sqrt{A_A}$ is similar when we consider restricted communities, with a linear initial growth, followed by a first plateau for intermediate academic ages, and a decrease to a lower plateau for longer academic ages: the observed plateaus are summarized on table~\ref{table1}, under the column "Plateau", with the larger value referring to the former plateau, and the smaller value to the latter plateau; the column "Sample" shows the number of physicists falling in each category.
We emphasize that the constant behaviour of the quantity $h\sqrt{A_A}$ over the large region of academic ages between 12 and 24 years suggests that indeed its plateau value could be used as a quality benchmark.

\begin{figure}[hbt]
\includegraphics*[width=8 cm]{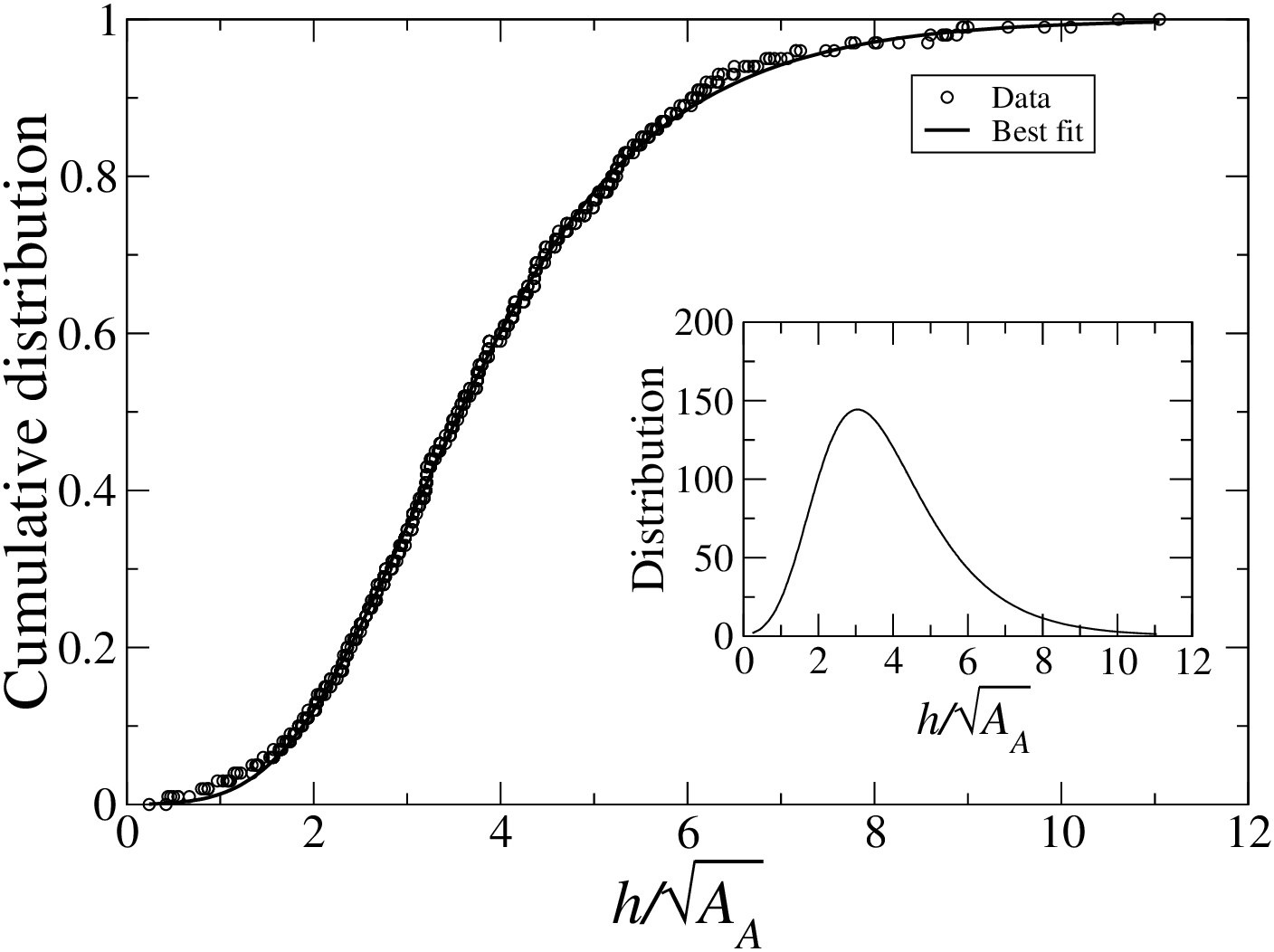}
\caption{The cumulative distribution of average $h$-index normalized to the square root of academic age $h/\sqrt{A_A}$, for individuals with academic age between 12 and 24 years (symbols). The solid line is a best fit using a Gompertz function (see text). The inset shows the distribution of the number of individuals with a given $h/\sqrt{A_A}$, obtained taking the derivative of the best fitting Gompertz function, scaled so that the area under the curve equals the number of researchers in this $A_A$ range (554).}
\label{fig4}
\end{figure}

\begin{figure}[hbt]
\includegraphics*[width=8 cm]{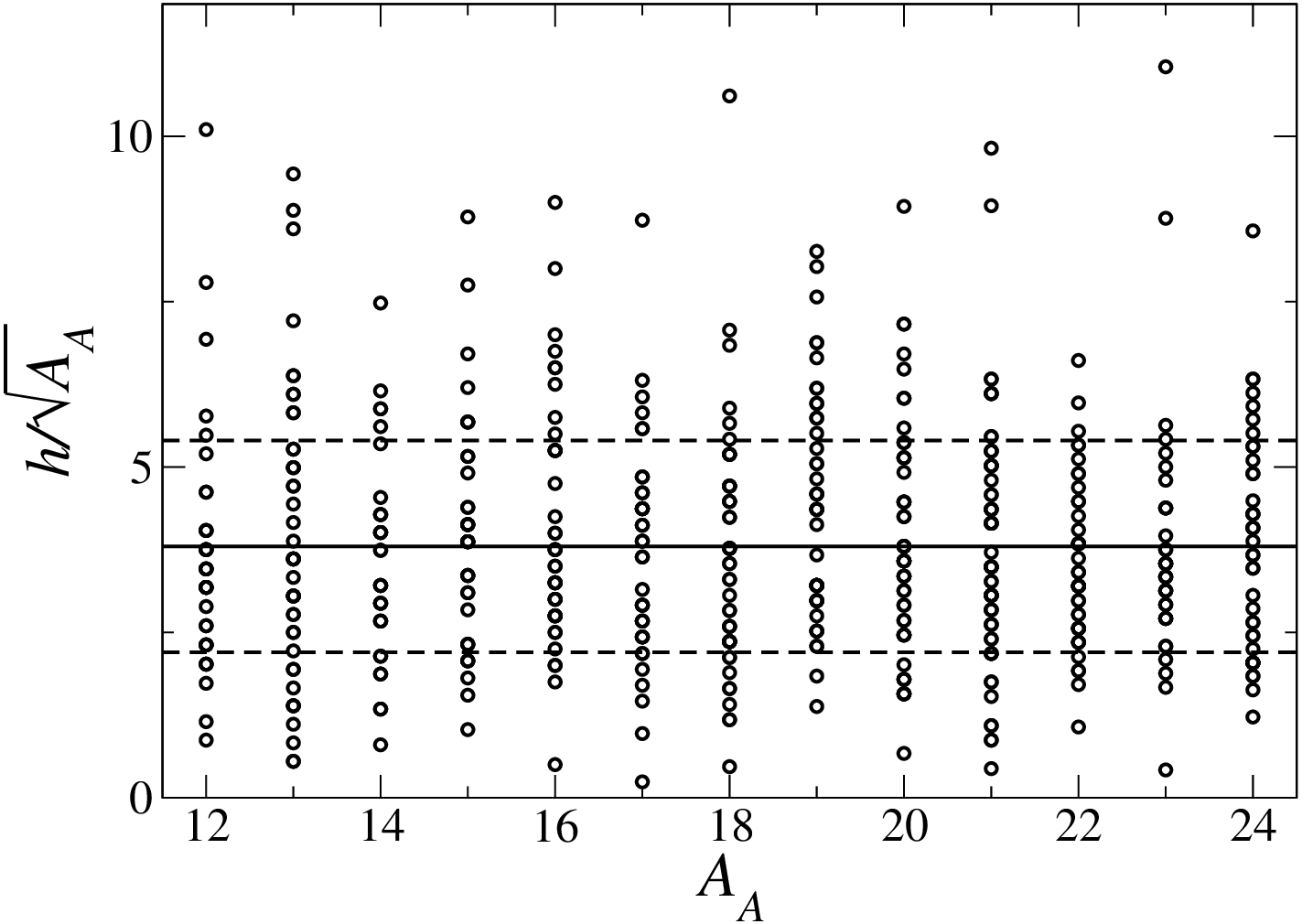}
\caption{An enlargement of the central region ($A_A$ in the range from 12 to 24 years) of Fig.~\protect{\ref{fig3}}. The symbols show $h\sqrt{A_A}$ for each individual; the solid and dashed lines mark the average and the one standard deviation values of  $h\sqrt{A_A}$ ($3.8\pm1.6$).}
\label{fig5}
\end{figure}

\section{The time scaled index $h\sqrt{A_A}$}

It is interesting to assess the statistical properties of the distribution of the index $h\sqrt{A_A}$. The main
result is shown in Fig.~\ref{fig4}: as customary in the presence of discrete distributions characterized by some fluctuations, we first studied the cumulative distribution of the quantity $h/\sqrt{A_A}$ (shown as small circles in the figure), for individuals with academic ages in the intermediate range. These data are very well described by the Gompertz function
$f(x) = \exp(-e^{-c(x-b)})$~\citep{Gompertz}, with $c$ and $b$ parameters quantifying the data. The fit to the data using this function is shown by the solid line, where the parameters turn out to be $c=0.71$ and $b=3.05$. The inset shows the derivative of the fitting function - in other words, the smoothed distribution of expected values for the number of individuals having a given value of $h/\sqrt{A_A}$. We notice that given the skewness of the derivative, the value of $b$ (which yields the position of the maximum of the derivative) is smaller than the computed average value (see table~\ref{table1}). On the other hand, the inverse of $c$ is a good indicator of the width of the derivative, and it follows that $1/c \approx 1.4$, which is pleasantly close to the standard deviation (1.6) we computed directly from the data. Finally, Fig.~\ref{fig5} shows an enlargement of the central region of Fig.~\ref{fig3}: the average value of $h/\sqrt{A_A}$ and the one standard deviation lines nicely interpolate the distribution of $h/\sqrt{A_A}$, for the whole range of academic ages considered.

\begin{figure}[hbt]
\includegraphics*[width=8 cm]{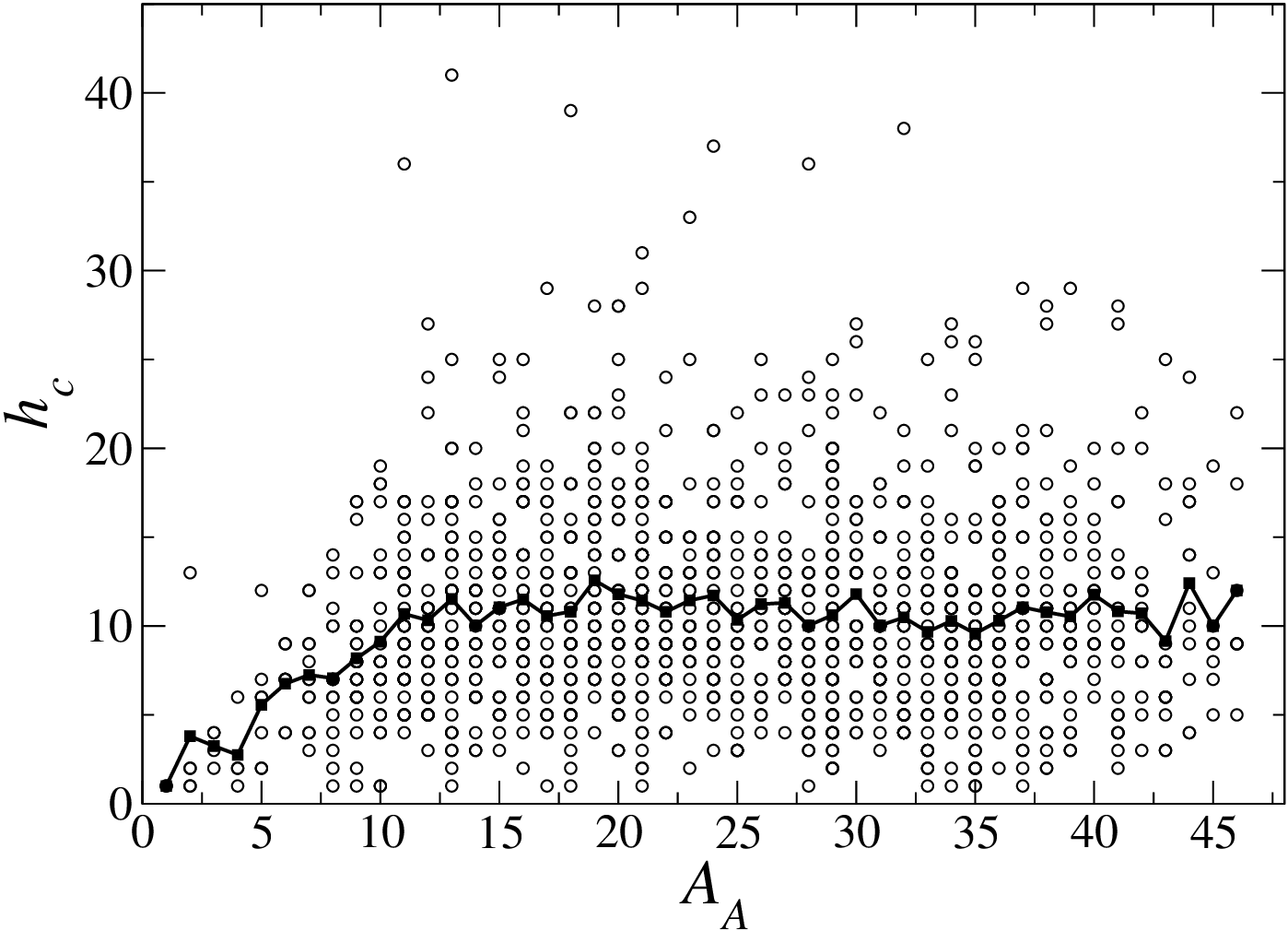}
\caption{The average $h_c$ vs the academic age $A_A$ (black squares with a broken line). Circles show the values of the $h_c$ of each researcher in our sample.}
\label{fig6}
\end{figure}

\begin{figure}[hbt]
\includegraphics*[width=8 cm]{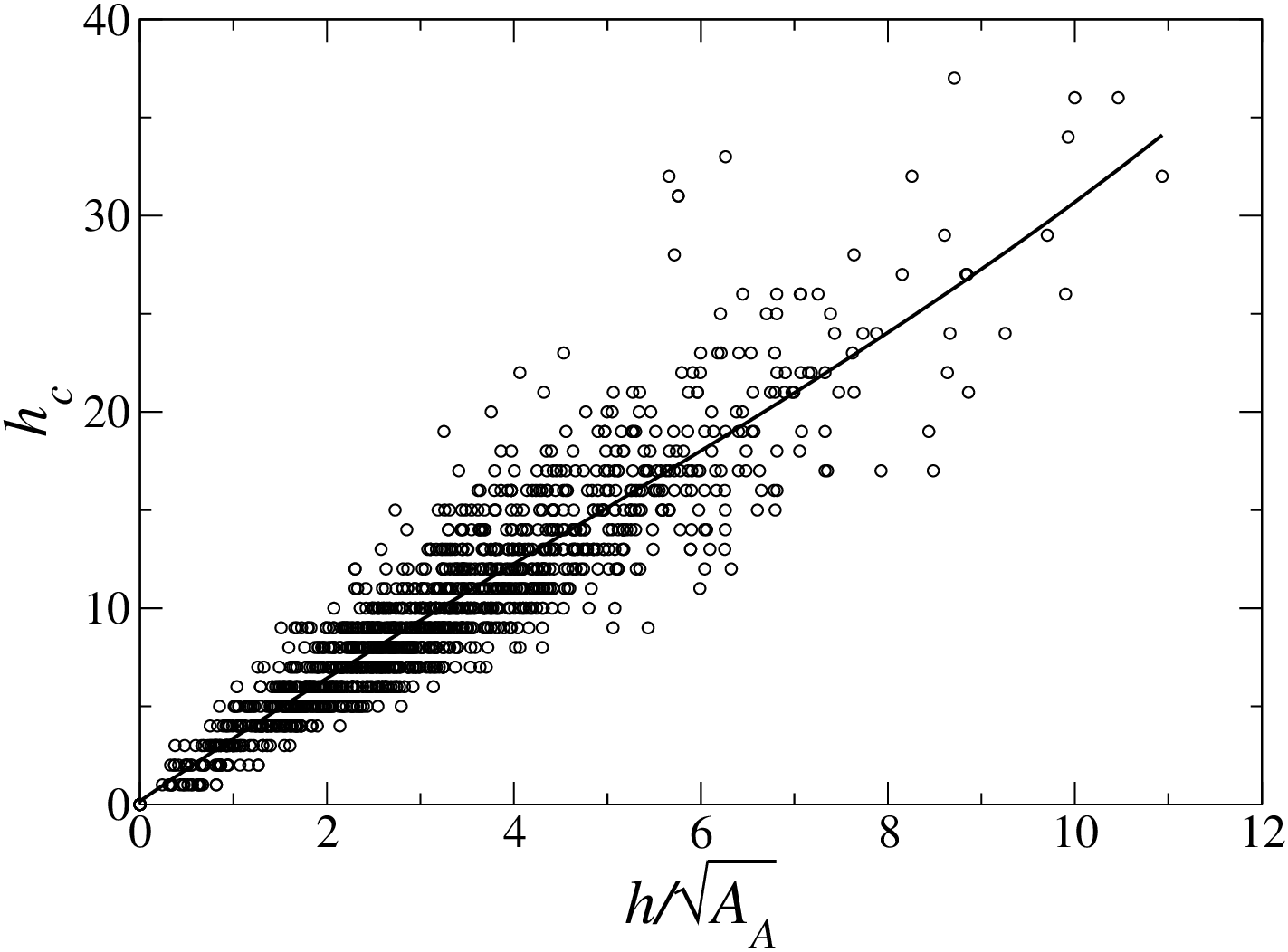}
\caption{Comparison between the contemporary $h$-index  $h_c$ and the index $h/\sqrt{A_A}$ (symbols). The solid line is a best fit using a cubic polynomial.}
\label{fig7}
\end{figure}

\section{The comparison with the contemporary $h$-index}

It is also interesting to compare the index $h\sqrt{A_A}$ to the contemporary $h$-index~\citep*{SKM:2006} ($h_c$ in the following), which has been introduced to assess individuals who have been scientifically active over widely different ranges of time. We recall that $h_c$ depends on the current year $y_n$, and it is evaluated by renormalizing the number of citations $n_i$ of the paper $i$, published in the year $y_i$, as $\tilde{n}_i = n_i \gamma (y_n-y_1+1)^{-\delta}$, and using the $\tilde{n}_i$ sequence to compute $h_c$, with the same algorithm used for the $h$-index. 

To carry out the comparison, we took the widely used values $\gamma=4$ and $\delta=1$. First, we plotted on Fig.~\ref{fig6} the contemporary $h$-index ($h_c$) as a function of the academic age. Circles are the $h_c$ index for each individual in our sample, and the black square joined by a broken solid line show the average $h_c$ for each $A_A$ class. It is very interesting to notice that, after an initial region where the average $h_c$ grows linearly, for $A_A$'s larger than 12 years the average $h_c$ remains constant, up to the largest $A_A$ present in our sample. 

The comparison between $h_c$ and $h\sqrt{A_A}$ is summarized in Fig.~\ref{fig7}, where we plotted $h_c$ versus the index $h/\sqrt{A_A}$, for all individuals in our sample. It is clear from the figure that the two indicators are proportional to each other, and this is confirmed by a best fit using a cubic polynomial, shown in the figure as a solid line, which appears indistinguishable from a straight line. The conclusion is that, at least for our sample, the index $h/\sqrt{A_A}$ appears to yield the same information provided by the contemporary $h$-index, and hence the two indexes are interchangeable (at least within a wide academic age range): we like to remark, however, that the evaluation of the quantity $h/\sqrt{A_A}$ appears to be easier than the evaluation of the contemporary $h$-index, and that the contemporary $h$ index requires two arbitrary parameters ($\gamma$ and $\delta$) which need to be introduced empirically. It will be matter for further work to assess whether the proportionally between these indexes is also observed when different values for the two parameters are taken.

\section{Conclusions}

We have produced evidence that the index $h/\sqrt{A_A}$, averaged over sufficiently large groups, is a sensible proxy for the contemporary $h$-index, and tends to stay constant in time in the interval between 12 and 24 years of research activity, which is the typical range for researchers to apply for permanent and/or higher positions. The plateau value the index $h/\sqrt{A_A}$ might therefore be used as a quality benchmark, even if its eminently statistical origin does not make it proper to employ it for any kind of ranking of individual researchers.

As for the numerical value of  the plateau, one must not forget that our analysis implied the aggregation of  widely different typologies of researchers, and therefore the numbers we obtained are weighted averages of the values corresponding to each homogeneous subgroup of researchers. This should not affect our general conclusions, since we have revealed a common trend, and the lack of homogeneity could, at most, obscure specific trends that are peculiar to a subgroup.

\section{Acknowledgements}

Discussions with S. Benedetto, A. Bonaccorsi and G. Parisi on the topic of this paper are warmly acknowledged.

\bibliographystyle{model5-names}

\end{document}